\newif \ifarxiv
\arxivtrue

\ifarxiv
\documentclass[%
aps,pre,
superscriptaddress,
 showpacs,showkeys
]{revtex4-1} 
\else
\documentclass[%
aps,pre,
twocolumn,superscriptaddress,
 showpacs,showkeys
]{revtex4-1} 
\fi

\usepackage{graphicx,dblfloatfix}
\usepackage{amsmath,amssymb}
\usepackage{mathtools}
\usepackage{url}
\usepackage{bm}
\usepackage{color}

\usepackage{xr}
\usepackage{graphicx,dblfloatfix}
\usepackage{amsmath,amssymb}
\usepackage{mathtools}
\usepackage{hyperref}
\usepackage{bm}
\usepackage{color}
\usepackage{graphicx}
\usepackage{tikz,pgf}
\usepackage{pgfplots}
\usetikzlibrary{calc}
\pgfplotsset{compat=newest}
\pgfplotsset{width=7cm,compat=1.8}
\usetikzlibrary{spy}
\usetikzlibrary{positioning}
\usetikzlibrary{decorations.text}
\usetikzlibrary{decorations.pathmorphing}
\usetikzlibrary{arrows.meta}
\usetikzlibrary{shapes.multipart}
\usetikzlibrary{fadings,shapes.arrows,shadows}   
\tikzset{>=latex}
\tikzset{arrowfill/.style={fill=jzred!20}}
\tikzset{arrowstyle/.style={font=\scriptsize,draw=jzred,arrowfill, single arrow,
single arrow head extend=.2cm,}}
\tikzstyle{blockjz} = [rectangle, align=center, minimum width={width("Microscopic picture")+.8cm}, minimum height = 2em,font=\sffamily\scriptsize]
  \newenvironment{customlegend}[1][]{%
        \begingroup
        \csname pgfplots@init@cleared@structures\endcsname
        \pgfplotsset{#1}%
    }{%
        \csname pgfplots@createlegend\endcsname
        \endgroup
    }%
    \def\addlegendimage{\csname pgfplots@addlegendimage\endcsname}
\newif\ifexportimage
\exportimagetrue
\newif\iftikzjz
\tikzjztrue 
\tikzjzfalse 
\newcommand{\pathname}{}
\iftikzjz
\usetikzlibrary{external}
\fi
\definecolor{jzorange}{RGB}{230, 159, 0}
\definecolor{jzskyblue}{RGB}{86, 180, 233}
\definecolor{jzgreen}{RGB}{0, 158, 115}
\definecolor{jzyellow}{RGB}{240, 228, 66}
\definecolor{jzblue}{RGB}{0, 114, 178}  
\definecolor{jzred}{RGB}{213, 94, 0} 

\renewcommand{\underline}[1]{#1}

\newcommand{\E}{\mathbb{E}}
\newcommand{\K}{{\mathcal K}}
\newcommand{\DS}{\frac{\delta S(z)}{\delta z}}
\newcommand{\DSrho}{\frac{\delta S(\rho)}{\delta \rho}}
\renewcommand{\S}{S}
\newcommand{\eps}{\epsilon}

\newcommand{\df}{\,\mathrm{d}}
\newcommand{\dd}{\mathrm{d}}
\renewcommand{\div}{\mathop{\mathrm{div}}}

\newcommand{\discr}{\mathrm{discr}}

\ifarxiv

\else

\fi

\begin{document}

\preprint{AIP/123-QED}

\title[]{How to find the evolution operator of dissipative PDEs from particle fluctuations?}

\author{Xiaoguai Li}
\affiliation{Department of Mechanical Engineering and Applied Mechanics, \\ University of Pennsylvania, Philadelphia, PA 19104, USA}
\author{Nicolas Dirr}
\email[]{dirrnp@cardiff.ac.uk}
\affiliation{School of Mathematics, Cardiff University, Cardiff CF24 4AG, UK}
\author{Peter Embacher}
\affiliation{Mathematics Institute, University of Warwick, Coventry CV4 7AL, UK}
\author{Johannes Zimmer}
\email[]{zimmer@maths.bath.ac.uk}
\affiliation{Department of Mathematical Sciences, University of Bath, Claverton Down, Bath BA2 7AY, UK}
\author{Celia Reina}
\email[]{creina@seas.upenn.edu}
\address{Department of Mechanical Engineering and Applied Mechanics, \\ University of Pennsylvania, Philadelphia, PA 19104, USA}

\date{\today}




\begin{abstract}
  \def\boldsymbol{} Dissipative processes abound in most areas of sciences and can often be abstractly written as $\boldsymbol
  \partial_t \boldsymbol z = \boldsymbol \K(\boldsymbol z) \boldsymbol\delta \boldsymbol \S(\boldsymbol z)/\boldsymbol\delta
  \boldsymbol z$, which is a gradient flow of the entropy $\boldsymbol\S$. Although various techniques have been developed to
  compute the entropy, the calculation of the operator $\boldsymbol \K$ from underlying particle models is a major long-standing
  challenge.  Here, we show that discretizations of diffusion operators $\boldsymbol \K$ can be numerically computed from particle
  fluctuations via an infinite-dimensional fluctuation-dissipation relation, provided the particles are in local equilibrium with
  Gaussian fluctuations.  A salient feature of the method is that $\boldsymbol \K$ can be fully pre-computed, enabling macroscopic
  simulations of arbitrary admissible initial data, without any need of further particle simulations. We test this coarse-graining
  procedure for a zero-range process in one space dimension and obtain an excellent agreement with the analytical solution for the
  macroscopic density evolution.  This example serves as a blueprint for a new multiscale paradigm, where full dissipative
  evolution equations --- and not only parameters --- can be numerically computed from particles.
\end{abstract}

\maketitle




\section{Introduction}

The art of modeling faces a deep gulf between macroscopic continuum models that are efficiently computable, although typically
riddled with phenomenology, and lower scale atomistic/particle simulations, which are of higher physical fidelity, yet often
exceedingly costly for real-life applications. Bridging this gulf has been a major research endeavor in many disciplines,
e.g.~mathematics, physics or chemistry, leading to numerous coarse-graining strategies. Some prominent examples to determine the
macroscopic evolution include methods based on projection operators~\cite{Ottinger2005a}, multiscale
techniques~\cite{Abdulle2012a}, equation-free methods~\cite{Kevrekidis2009a} and information-theoretic
strategies~\cite{Machta2013a}; see also~\cite{Givon2004a} for a review of mathematical approaches. However, few rigorous results
are available, notably from hydrodynamic limit theory~\cite{Kipnis1999a} or strategies as in~\cite{Bodineau2016a}. This topic
belongs to the vast field of macroscopic evolution discovery, for example via dimension
reduction~\cite{Stephens2011a,Daniels2015a} and machine learning techniques~\cite{Rudy2017a,Raissi2018a}, just to mention a few
recent approaches.

\begin{figure}[b!]
  \centering
  \newcommand{\tikzpicturename}{fig-sketch-idea}
  \iftikzjz \tikzsetnextfilename{\tikzpicturename}
  \input{\pathname\tikzpicturename.tikz}
  \else\includegraphics{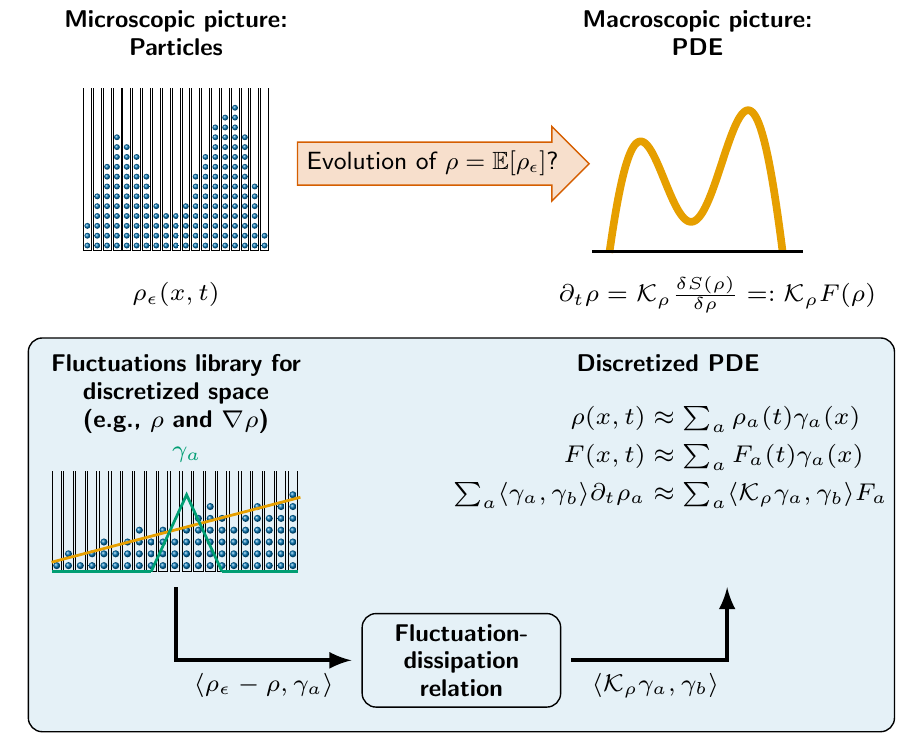}\fi  
  \caption{Sketch of the proposed computational strategy to determine the dissipative operator $\K$ in discretized form, using
      basis functions $\{\gamma(x)\}$, for the specific case of a density field $\rho$.}
    \label{fig:diagram}
\end{figure}

Here, we propose a fundamentally distinct strategy to coarse-grain the entire evolution for dissipative particle processes.  The
methodology, which is sketched in Fig.~\ref{fig:diagram} for a mass diffusion problem, numerically computes the macroscopic
evolution operator from the underlying field fluctuations. It presents three main attractive features. Firstly, the operator can
be completely pre-computed, enabling continuum simulations that do not require concurrent particle calculations. Secondly, the
method presented here computes the operator fully---not only parameters---and does not require a library of pre-existing
operators. Thirdly, the computation of the operator only requires the fluctuations of the macroscopic fields, which could in
principle be determined experimentally. The first feature distinguishes this approach from existing multi-scale techniques and
equation-free methods, while the second one differentiates it from machine learning approaches.

A central element of the strategy is a novel infinite-dimensional fluctuation-dissipation statement, which generalises existing
finite-dimensional results. The latter have been widely used to determine transport coefficients (i.e., parameters in otherwise
fully specified operators)~\cite{Green1954b,Kubo1966a,Marconi2008a,Embacher2018a}. In contrast, the new approach allows us to
infer the evolution operator (including parameters therein) by probing particle fluctuations systematically.

As a first step in this direction, we focus here on purely diffusive systems, and demonstrate the approach for a zero range
process.

\section{Outline of the approach}

Underlying the proposed strategy is the assumption that macroscopic dissipative evolutions in `thermodynamic' form can be written
as
\begin{equation}
   \label{eq:GF}
   \partial_t z= \K(z) \DS =: \K_z \DS,
\end{equation}
where $z=z(x,t)$ is the field of interest, $\S$ is the entropy of the system, $\DS$ its variational derivative and $\K_z$ is a
symmetric positive semi-definite linear operator; we write $\K_z$ to emphasize that $\K$ depends on $z$.  On the level of
nonequilibrium thermodynamics/mechanics,~\eqref{eq:GF} describes a wide range of evolution equations.

The specific question addressed in this paper is the following: given a particle model that leads, in a suitable scaling limit of
infinitely many particles, to an equation of the form~\eqref{eq:GF}, how can we determine the operator $\K_z$ purely from the
observation of finitely many particles? We focus on $\K_z$ since the computation of $\DS$ can, in many situations, be accomplished
by free energy computations~\cite{Lelievre2010a}.

The key observation is that the evolution of a large, yet \emph{finite} number of particles, can often be formally described by a
stochastic partial differential equation of the form
\begin{equation}
  \label{eq:stoch-GF}
  \partial_t z_{\eps} =\K_{z_{\epsilon}}\frac{\delta \S(z_{\epsilon})}{\delta z_{\epsilon}}+\sqrt{\epsilon}\sqrt{2\K_{z_{\epsilon}}}\dot{W}_{x,t},
\end{equation}
where $\dot{W}_{x,t}$ is a space-time white noise, $\E\left[\dot W_{x,t} \dot W_{y,s}\right] = \delta(x-y) \delta(s-t)$, and the
equation is to be interpreted in a weak formulation. This is the fluctuating hydrodynamics equation associated with~\eqref{eq:GF},
for diffusive systems~\cite[Section 6]{Eyink1990a},~\cite[Section 4]{Eyink1996a}, or those described by an additive noise. It
encodes an infinite-dimensional fluctuation-dissipation relation $\sigma_z \sigma_z^*=2 \epsilon \K_z$, with the fluctuation
operator $\sigma_z = \sqrt{\epsilon}\sqrt{2\K_{z_{\epsilon}}}$ acting on the noise. Its finite dimensional counterpart has long
being used to extract transport coefficients. See, for example, the monographs~\cite{Stratonovich1992a,Zwanzig2001a,Klages2013a},
and the articles~\cite[Section~3]{Eyink1996a},~\cite{Maes1999a} for fluctuation-dissipation relations.

As a simple example of~\eqref{eq:stoch-GF}, the evolution of $N$ random walkers $X_i$ on a lattice described by $\rho_{\epsilon}
:= \frac 1 N \sum_{i=1}^N \delta_{X_i}$ is given by an equation of Dean type~\cite{Dean1996a} ($N =C/\epsilon$, $\epsilon$ being
the individual lattice site volume and $C$ a constant)
\begin{equation}
  \label{eq:dean}
  \partial_t \rho_{\epsilon} = \div (D \nabla \rho_{\epsilon}) + \sqrt{\epsilon} \div(\sqrt{2 D\rho_{\epsilon}} \dot{W}_{x,t}).
\end{equation}
This equation is of the form~\eqref{eq:stoch-GF}, with ${\S(\rho) = -\int \rho \log(\rho) \df x}$ being the Boltzmann entropy in
dimensionless units and $\K_{\rho}$ the operator ${\K(\rho) \xi = -\div(D\rho \nabla \xi)}$; see~\cite{Reina2015a} for the
calculation of $\sqrt{\K_{\rho}}$ in this case (note that Dean derives an equation for $\sum_{i=1}^N \delta_{X_i} = N
\rho_\eps(x)$, which is why the noise in~\cite{Dean1996a} differs from the one in~\eqref{eq:dean} by a factor of $\sqrt{N}$; this
difference in scaling is crucial, see~\cite[Section~6]{Eyink1990a}). There is a wide range of applications of~\eqref{eq:stoch-GF}
and its extension to account for reversible phenomena, for example, nonequilibrium bacterial
dynamics~\cite[Eq.~(71)]{Thompson2011a}, nucleation theory~\cite[Eqs.~(8) and~(20)]{Lutsko2012a} and liquid film
theory~\cite[Eq.~(19)]{Grun2006a}; see~\cite{Duran-Olivencia2017a} for connections to dynamic density functional theory. We remark
that Macroscopic Fluctuation Theory~\cite{Bertini2015a} is also a fluctuation-based theory to describe the macroscopic evolution
of interacting particle systems; there, the evolution of the macroscopic density and flux are determined via Large Deviation
Theory, leading to an equation of the form ${\partial_t \rho = \div(D(\rho) \nabla \rho - \chi(\rho) E(t)}$),
see~\cite[(2.3)]{Bertini2015a}. In this article, $E\equiv 0$; the noise of~\eqref{eq:dean} is the one associated with Large
Deviation Theory~\cite{Jack2014a}, though we do not use this fact here.

\section{Numerical procedure}

We now show that the fluctuation-dissipation relation in~\eqref{eq:stoch-GF} can be harnessed to numerically compute a discretized
version of the operator $\K_z$ from particle data. More specifically, we consider an approximation of the macroscopic field $z$
and its associated thermodynamic force $F:=\delta S/\delta z$ of the form $z(x,t) \approx \sum_a z_{a}(t) \gamma_a(x)$ and $F(x,t)
\approx \sum_a F_{a}(t) \gamma_a(x)$, where $\{\gamma_a\}$ is a suitable basis of functions. The weak form of the evolution
equation~\eqref{eq:GF} then reads
\begin{equation} 
  \label{Eq:WeakFormPDE}
  \sum_a \langle \gamma_a, \gamma_b \rangle \partial_t z_a = \sum_a \langle \K_z\gamma_a,\gamma_b \rangle F_a,
  \text{ for all } b,
\end{equation}
where the matrix $\langle \K_z\gamma_a,\gamma_b \rangle$ represents a discretization of the unknown operator. In analogy to the
quadratic variation formula for the stochastic ODE $\df X= f \df t + \sqrt{\sigma} \df W_t$ in dimension $n$, where
\begin{equation*}
 \sigma= \lim_{h\searrow 0}\frac{1}{h n}\E\left[\left[X(t_0+h)-X(t_0)\right]^2\right],
\end{equation*}
the element $\langle \K_z\gamma_a,\gamma_b \rangle$ can be related to the covariation of the rescaled local fluctuations as
\ifarxiv
\begin{equation}
 \langle \K_z\gamma_a,\gamma_b \rangle
  =\lim_{h \searrow 0} \frac{1}{2h}\mathbb{E}\Big[ \left(Y_{\gamma_a}(t_0+h)- Y_{\gamma_a}(t_0)\right) \Bigr.  
   \Bigl. \cdot \left(Y_{\gamma_b}(t_0+h)- Y_{\gamma_b}(t_0) \right)\Big],
  \label{Eq:DiscreteMetric}
\end{equation}
\else
\begin{multline}
 \langle \K_z\gamma_a,\gamma_b \rangle
  =\lim_{h \searrow 0} \frac{1}{2h}\mathbb{E}\Big[ \left(Y_{\gamma_a}(t_0+h)- Y_{\gamma_a}(t_0)\right) \Bigr.  
  \\  
   \Bigl. \cdot \left(Y_{\gamma_b}(t_0+h)- Y_{\gamma_b}(t_0) \right)\Big],
  \label{Eq:DiscreteMetric}
\end{multline}
\fi
where $Y_{\gamma}$ is the limit of $\langle z_{\epsilon}-z,\gamma \rangle /\sqrt{\epsilon}$ and $z=\E[z_{\epsilon}]$. The proof of
this statement can be found in Appendix~\ref{sec:covariance}. It uses mathematical arguments similar to~\cite{Embacher2018a}, but
extends the result considerably: now the entire operator can be characterized, while in~\cite{Embacher2018a} only one parameter
(the diffusivity) could be extracted for an otherwise fully prescribed operator.

\begin{figure}[t]
  \centering
  \newcommand{\tikzpicturename}{fig-test-functions}
  \iftikzjz \tikzsetnextfilename{\tikzpicturename}
  \input{\pathname\tikzpicturename.tikz}
  \else\includegraphics{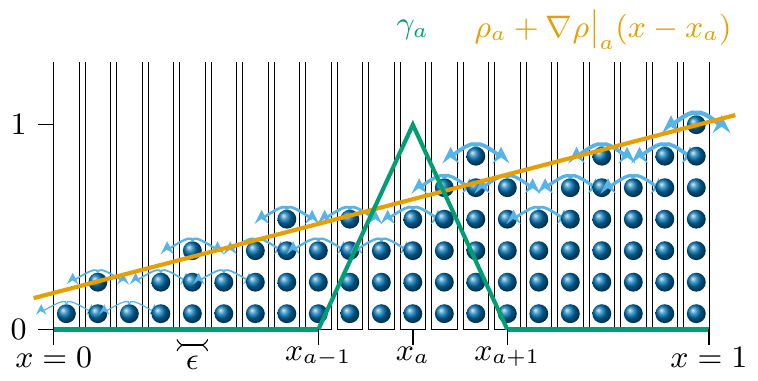}\fi
  \caption{Sketch of a profile $\rho_a + \nabla \rho|_a (x - x_a)$ simulated to measure density fluctuations, and a basis function
    $\gamma_a$. The jump rate for each lattice site increases with the number of particles, as indicated by the thickness of the
    arrows. }
  \label{Fig:ProfilesSimulated}
\end{figure}

Relation \eqref{Eq:DiscreteMetric} enables the calculation of the discretized operator $\K_z$ from particle fluctuations. However,
the result is, in general, profile dependent, as indicated by the subscript $z$ in $\K_z$, and thus the argument $z$ is still
infinite-dimensional. This is another key difference to~\cite{Embacher2018a}, where the diffusivity depends on a scalar density
value. Therefore additional arguments are required to precompute the operator. Namely, we consider functions $\gamma$ with local
support and assume that $\K_z$ is a local and regular operator, such that a Taylor approximation in $z$ can be employed (these
assumptions are satisfied for a wide range of operators and suitable choices of $z$). Then, a numerical approximation of the
left-hand side of~\eqref{Eq:DiscreteMetric} is
\begin{equation}
  \label{Eq:ApproxDiscreteOperator}
  \langle \K_z\gamma_a,\gamma_b \rangle \approx    \langle
  \K_{\left(z_a+ \nabla z|_a (x-x_a) +\ldots\right)}\gamma_a,\gamma_b \rangle 
\end{equation}
for sufficiently high order of the Taylor expansion and suitable basis functions; here $z_a = z(x_a)$, where $x_a$ the mid-point
of $\gamma_a$, see Fig~\ref{Fig:ProfilesSimulated}, and $\nabla$ is the spatial gradient. We remark that the assumption of
locality of the operator can be numerically probed, by evaluating $\langle \K_z \gamma_a, \gamma_b \rangle$ for functions
$\gamma_a$ and $\gamma_b$ with non-overlapping support.

In practice, the calculation of $\langle \K_{\left(z_a+ \nabla z|_a (x-x_a)+ \ldots \right)}\gamma_a,\gamma_b \rangle$
in~\eqref{Eq:ApproxDiscreteOperator} as a function of $z_a, \nabla z|_a, \ldots$ is implemented for a discretized space
$V_\discr$, i.e., for finitely many values of $z_a, \nabla z|_a, \ldots$ within a prescribed range (see
Fig.~\ref{Fig:ProfilesSimulated} for $z\coloneqq\rho$, and the space $V$ given by $\rho$ and $\nabla \rho$). This is obtained
via~\eqref{Eq:DiscreteMetric} from particle data (finite $\epsilon$), for small but finite $h$, and expectations are approximated
as averages over $R$ realizations. The resulting discrete operator is then interpolated in $V$ to perform continuum simulations
with~\eqref{Eq:WeakFormPDE} and~\eqref{Eq:ApproxDiscreteOperator}, for arbitrary initial conditions and boundary data (here
periodic or Dirichlet boundary conditions).  Although the pre-calculation of $\K_z$ can be laborious, requiring a considerable
number of simulations (of the order of $R\ \times$ size of the discrete space $V_\discr$), these are trivially parallelizable and
only executed over a small time interval. Moreover, once the operator is computed, the macroscopic simulations can be run without
any further particle simulations.

\section{Computational results}

We now demonstrate the applicability of this coarse-graining strategy with an illustrative example, where the the analytical
solution is known, and the errors may thus be quantified. Specifically, we consider a symmetric zero-range process (ZRP) in a
one-dimensional lattice. Here, particles jump with a rate $g(k)=k^2$, where $k$ is the occupation number of the specific site, see
Fig.~\ref{Fig:ProfilesSimulated}. This particle process can be efficiently simulated using a Lattice Kinetic Monte Carlo approach,
and, in the limit of infinite number of particles ($\epsilon \rightarrow 0$), the density profile evolves according to the PDE
\begin{equation}
  \label{eq:ZRP-thermo}
  \begin{split}
    &\partial_t \rho=-\div \left(m(\rho) \nabla \DSrho \right), \quad \text{with}\\
    & \frac{\delta S}{\delta \rho}=-\log \left(2 m(\rho) \right), \ \text{and } \rho(m)=\sqrt{2m} \frac{I_1\left(2\sqrt{2m}
      \right)}{I_0\left(2\sqrt{2m} \right)}, \\
  \end{split}
\end{equation}
where $I_i$ are the modified Bessel functions of the first kind; see~\cite{Grosskinsky2003a,Embacher2018a} for the derivation.

\begin{figure}[t]
 \centering
     { 
       \pgfplotsset{
         compat=1.11,
         legend image code/.code={
           \draw[ultra thick, mark repeat=2,mark phase=2]
           plot coordinates {
             (0cm,0cm)
             (0.04cm,0cm)        
             (0.08cm,0cm)         
           };%
         }
       }
         \newcommand{\tikzpicturename}{fig-matrix-entries-constraint-V-disc}
         \iftikzjz \tikzsetnextfilename{\tikzpicturename}
         \input{\pathname\tikzpicturename.tikz}
         \else\includegraphics[scale=.57]{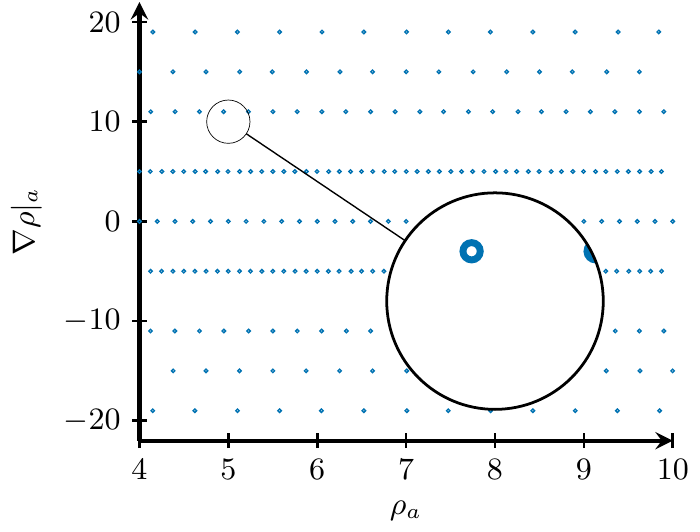}\fi
         \renewcommand{\tikzpicturename}{fig-matrix-entries-constraint-a-Equal-b}
         \iftikzjz \tikzsetnextfilename{\tikzpicturename}
         \input{\pathname\tikzpicturename.tikz}
         \else\includegraphics[scale=.57]{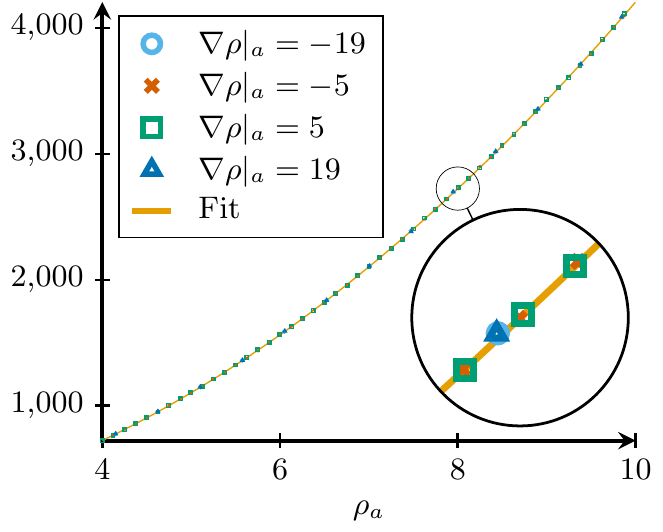}\fi
         \ifarxiv \\ \fi
         \renewcommand{\tikzpicturename}{fig-matrix-entries-constraint-a-Equal-bMinus1}
         \iftikzjz \tikzsetnextfilename{\tikzpicturename}
         \input{\pathname\tikzpicturename.tikz}
         \else\includegraphics[scale=.57]{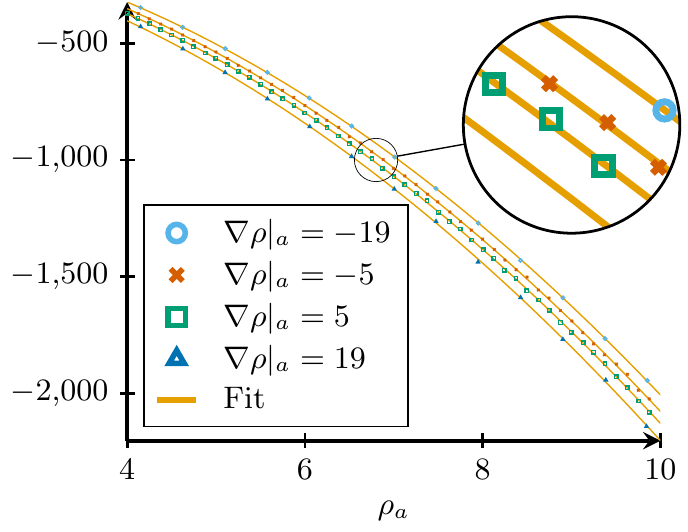}\fi
         \renewcommand{\tikzpicturename}{fig-matrix-entries-constraint-a-Equal-bPlus1}
         \iftikzjz \tikzsetnextfilename{\tikzpicturename}
         \input{\pathname\tikzpicturename.tikz}
         \else\includegraphics[scale=.57]{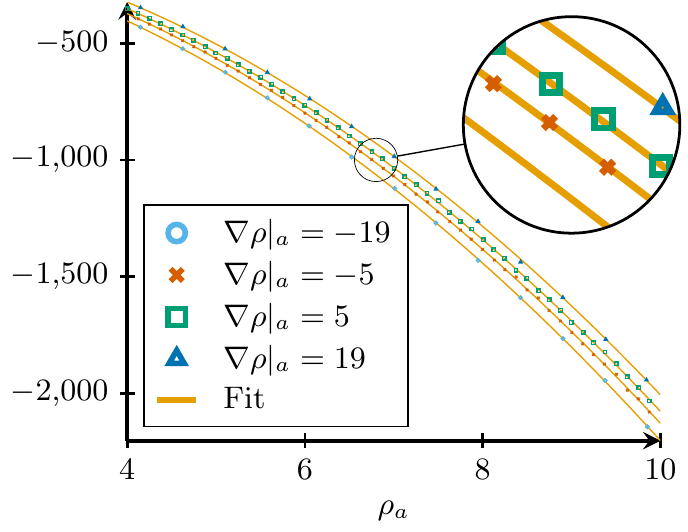}\fi
     } 
     \caption{Top left: Discrete set $V_\discr$ of pairs $(\rho_a, \nabla \rho|_a)$ used to evaluate the discretized operator
       $\K_\rho$. Remaining plots: Shown are the matrix entries $(b,a)$ of
       $\langle \K_{\left(\rho_a+ \nabla \rho|_a ( x-x_a)\right)}\gamma_a,\gamma_b \rangle$ as function of $\rho_a$ and
       $\nabla \rho|_a$ for a symmetric zero-range process. The plots are for $a=b$ (top right), $a=b-1$ (bottom left) and $a=b+1$
       (bottom right). For $x$, $x_a$ and the profiles $\gamma_a$ see Fig.~\ref{Fig:ProfilesSimulated} in the main text.}
   \label{Fig:MetricEntries-constraint}
\end{figure}

We choose linear finite element shape functions satisfying $\gamma_a(x_b)=\delta_{ab}$ (see Fig.~\ref{Fig:ProfilesSimulated}),
which lead to a tri-diagonal matrix for $\langle \K_\rho\gamma_a,\gamma_b \rangle$, and a linear approximation for $\rho$ in
${\K}_{\rho}$. The discretized evolution equation for the density then reads (for all $b$)
\begin{equation}
  \label{Eq:ApproxDiscreteOperator_rho}
    \hspace{-0.2cm} \sum_{a} \langle \gamma_a, \gamma_b \rangle \partial_t \rho_a
  \approx \hspace{-0.4cm} \sum_{a\in\{b-1,b,b+1\}}
  \hspace{-0.6cm} \langle \K_{\left(\rho_a+ \nabla \rho|_a ( x-x_a)\right)}\gamma_a,\gamma_b \rangle F_a;
\end{equation}
we remark that this choice is an assumption made \emph{a priori} on $\K_\rho$ and that it can in principle be generalized to
higher-order elements and Taylor approximations. With these approximations, the discretized operator can be tabulated by means of
three components (${a\in\{b-1,b,b+1\}}$) which depend on $\rho$ and $\nabla \rho$, and which can be computed for a discrete space
$V_\discr$ from particle simulations. The probed space $V_\discr$ and the three non-zero entries (main, super- and sub-diagonals)
of $\langle \K_{\left(\rho_a+ \nabla \rho|_a ( x-x_a)\right)}\gamma_a,\gamma_b \rangle$ are plotted in
Fig.~\ref{Fig:MetricEntries-constraint}, together with a polynomial fit that ensures mass conservation (i.e., the sum of the three
entries being equal to zero); see Appendix~\ref{sec:Comp-details} for further details on these calculations.

\begin{figure}[t]
  \centering
  \newcommand{\tikzpicturename}{fig-results-constraint-periodic}
  \iftikzjz \tikzsetnextfilename{\tikzpicturename}
  \input{\pathname\tikzpicturename.tikz}
  \else\includegraphics{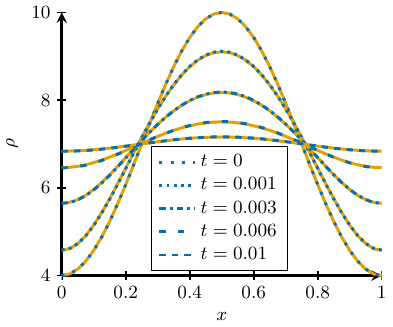}\fi
  \renewcommand{\tikzpicturename}{fig-results-constraint-Dirichlet}
  \iftikzjz \tikzsetnextfilename{\tikzpicturename}
  \input{\pathname\tikzpicturename.tikz}
  \else\includegraphics{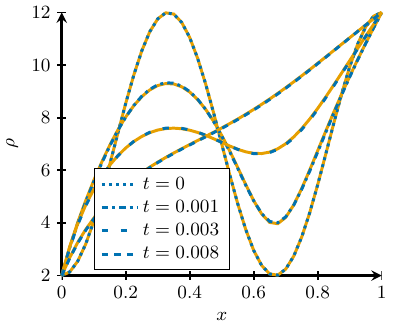}\fi
  \caption{Comparisons between the particle-based solution (blue) and the solution to the PDE from~\eqref{eq:ZRP-thermo} (orange)
    for different initial and boundary conditions. Left: Periodic boundary conditions. Right: Inhomogeneous Dirichlet
    data.}  \label{Fig:PDE_results-constraint}
\end{figure}

The resulting fitted operator in $V$ can then be utilized to compute the continuum evolution for arbitrary initial profiles, via
\eqref{Eq:ApproxDiscreteOperator_rho}. We here use the analytical thermodynamic force, cf.~\eqref{eq:ZRP-thermo}, to probe the
accuracy of the operator. Figure~\ref{Fig:PDE_results-constraint} shows two of such evolutions: the left figure depicts the time
progression of a cosine initial profile with periodic boundary conditions, whose density and gradient lie within the bounds of the
probed region in $V$, while the right figure considers a non-symmetric initial density with fixed Dirichlet data extrapolating
beyond this region. The full temporal evolution in movie format can be found as Movie~1 and Movie~2 in \ifarxiv
\cite{Li2018a-movies}. \else the Supplemental Material. \fi The results show an outstanding agreement between the
particle-informed evolution and the solution of the PDE given explicitly by~\eqref{eq:ZRP-thermo}, with an identical
spatio-temporal discretization scheme. A similar study without the mass conservation constraint is for comparison given in
Appendix~\ref{sec:Comp-details}. As it is there observed, results of good accuracy require an increase of the size of $V_\discr$
by more than an order of magnitude.

We further remark that the tabulated discrete operator can provide insight in its differential form, at least for simple cases.
In particular, for the zero-range process studied, whose infinite particle limit satisfies the PDE ${\partial_t \rho = - \div
  \left( m(\rho) \nabla F \right)= - m \Delta F - \nabla m \nabla F}$, see~\eqref{eq:ZRP-thermo}, the discrete operator reveals a
structure of the form
\begin{equation}
  \begin{split}
    \langle \K_{\left(\rho_a + \nabla \rho |_a(x-x_a) \right)}\gamma_a,\gamma_b\rangle& =: K_{ba} (\rho_a,\nabla \rho|_a)\\
    &=K_{ba}^{(1)}(\rho_a)+ K_{ba}^{(2)}(\rho_a, \nabla \rho|_a),
  \end{split}
\end{equation}
where the ratio of the coefficients result in a stencil of $K^{(1)}_{ba}$ equal to $-1, 1.999936, -0.999936$ (i.e., these are the
proportionality factors for the entries ${a=\{b-1,b,b+1\} }$ with $b$ fixed), which may be identified with the Laplacian; and a
stencil of $K^{(2)}$ equal to $-1, 0.003763, 0.996237$, which corresponds to the gradient. These stencils are obtained from the
constrained fitting method, noting that the discretized operator is symmetric up to higher order terms. The structure of the
right-hand side of the evolution equation is thus recovered.

This is, to the best of our knowledge, the first time that a dissipative PDE has been numerically recovered from particles using a
physics-based approach, in this case, an infinite-dimensional fluctuation-dissipation relation. We note that the required
assumptions are only threefold: a particle process which, for finitely many particles, is described by~\eqref{eq:stoch-GF}, and is
in local equilibrium with Gaussian fluctuations (see for example~\cite{Landim2002a} and~\cite[Section IV.B]{Dirr2016a} for a
precise mathematical formulation).

The example studied here can serve as a blueprint for the study of a wider spectrum of dissipative phenomena, with applications
ranging from two-phase flow over chemotaxis in heterogeneous environment to protein diffusion in membranes. Natural extensions are
also molecular dynamics simulations of Langevin type and lattice gas models. Such investigations should be complemented by a
rigorous numerical analysis of convergence and rates. These questions are beyond the scope of the present study and will be the
subject of future investigations.

\ifarxiv \paragraph{Acknowledgements}
\else \section*{Acknowledgements}
 \fi
All the authors thank the Leverhulme Trust for its support via grant no. RPG-2013-261. J.Z.\,gratefully acknowledges
  funding by a Royal Society Wolfson Research Merit Award. P.E.\,thanks Cardiff University through the International Collaboration
  Seedcorn Fund to visit C.R.; X.L.\,and C.R.\,thank the University Research Foundation Grant from the University of
  Pennsylvania.

\appendix

\section{Computational details for the zero-range process}
\label{sec:Comp-details}

In this section, we provide more information on the particle and continuum simulations for the zero-range process and compare the
coarse-grained results where mass conservation is incorporated with results where this constraint is not imposed.

For the process considered, we evaluate the discretized operator $\langle \K_{\left(\rho_a+ \nabla \rho|_a (x-x_a)
  \right)}\gamma_a,\gamma_b \rangle$ using simulations with flat profiles with $\rho_a \in [4:0.1:10]$, and affine profiles with
${\nabla \rho|_a \in \pm[5,11,15,19]}$, see Fig.~\ref{Fig:ProfilesSimulated} For each of these profiles, $R=800\,000$ realizations
are performed over the unit interval as computational domain, with $5000$ lattice sites (${\epsilon=1/5000}$). The system is first
evolved over a time interval $t_{0}-t_{\text{ini}}=4.004\times 10^{-6}$ to reach local equilibrium, and subsequently over a time
interval $h=4\times 10^{-11}$, which is used for the calculation of the expectations. In practice, the equilibration time interval
is much larger than $h$, yet macroscopically small, leading to negligible changes of the macroscopic profile.  To save
computational time, a method described in~\cite{Embacher2018a} is used to generate the multiple realizations. In addition, 40
equally spaced shape functions $\gamma_a$ are considered. This results in $125$ lattice sites within the support of each function
$\gamma_a$ that is fully contained in the computational domain. The resulting values of the matrix entries $\langle
\K_{\left(\rho_a+ \nabla \rho|_a (x-x_a) \right)}\gamma_a,\gamma_b \rangle$ evaluated at specific points $\left(\rho_a, \nabla
\rho|_a\right)$ in $V_\discr$ are then extended to $V$, via a suitable interpolation scheme. A na\"\i ve approach with an
independent interpolation of the individual matrix components will in general, however, not lead to a mass preserving scheme, in
contrast to what the particle process implies. Therefore, for the results of the main text, we imposed mass conservation in the
fitting process by ensuring that the entries in each column sum up to $0$. This is achieved using a least square fit with second
order polynomials in $(\rho_a, \nabla \rho|_a)$ for each of the three matrix entries, the sum of which we enforce to vanish
identically. Figure~\ref{Fig:MetricEntries-constraint} shows the chosen discrete set $V_\discr$ and resulting entries for the
operator, together with the polynomial fit. These were used to obtain Fig.~\ref{Fig:PDE_results-constraint}, where all simulations
employed an explicit time discretization scheme.

\begin{figure}[t]
  \centering
  \newcommand{\tikzpicturename}{fig-results-no-constraint-periodic}
  \iftikzjz \tikzsetnextfilename{\tikzpicturename}
  \input{\pathname\tikzpicturename.tikz}
  \else\includegraphics{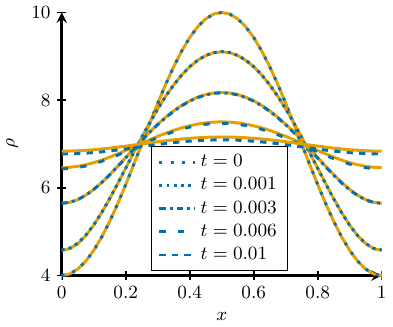}\fi
  \renewcommand{\tikzpicturename}{fig-results-no-constraint-Dirichlet}
  \iftikzjz \tikzsetnextfilename{\tikzpicturename}
  \input{\pathname\tikzpicturename.tikz}
  \else\includegraphics{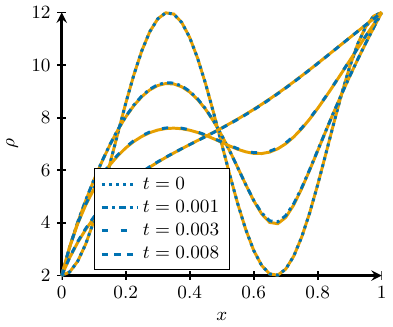}\fi
  \caption{Comparisons between the particle-based solution (blue) and the solution of the PDE from~\eqref{eq:ZRP-thermo} (orange)
    for different initial and boundary conditions, not imposing mass conservation on the discretized operator. Left: Periodic
    boundary conditions. Right: Inhomogeneous Dirichlet data.}
  \label{Fig:PDE_results}
\end{figure}

\begin{figure}[t]
 \centering
     { 
       \pgfplotsset{
         compat=1.11,
         legend image code/.code={
           \draw[ultra thick, mark repeat=2,mark phase=2]
           plot coordinates {
             (0cm,0cm)
             (0.04cm,0cm)        
             (0.08cm,0cm)         
           };%
         }
       }
         \newcommand{\tikzpicturename}{fig-matrix-entries-no-constraint-V-disc}
         \iftikzjz \tikzsetnextfilename{\tikzpicturename}
         \input{\pathname\tikzpicturename.tikz}
         \else\includegraphics[scale=.57]{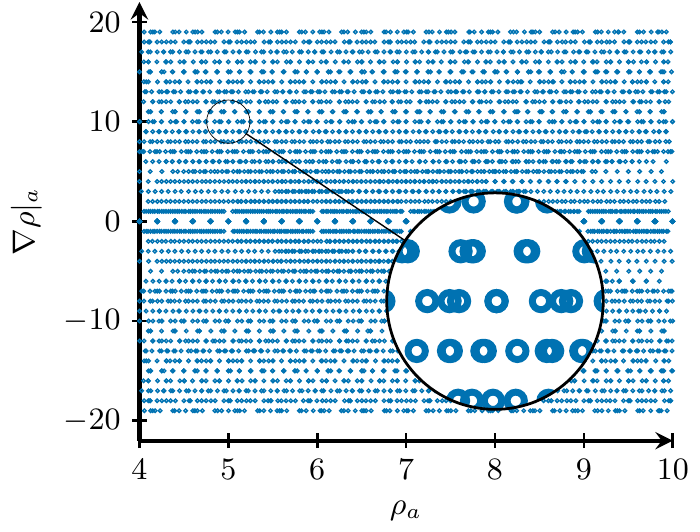}\fi
         \renewcommand{\tikzpicturename}{fig-matrix-entries-no-constraint-a-Equal-b}
         \iftikzjz \tikzsetnextfilename{\tikzpicturename}
         \input{\pathname\tikzpicturename.tikz}
         \else\includegraphics[scale=.57]{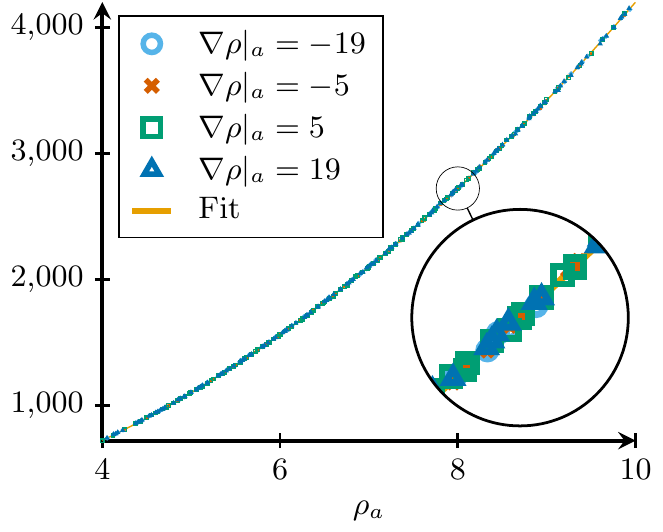}\fi
         \ifarxiv \\ \fi
         \renewcommand{\tikzpicturename}{fig-matrix-entries-no-constraint-a-Equal-bMinus1}
         \iftikzjz \tikzsetnextfilename{\tikzpicturename}
         \input{\pathname\tikzpicturename.tikz}
         \else\includegraphics[scale=.57]{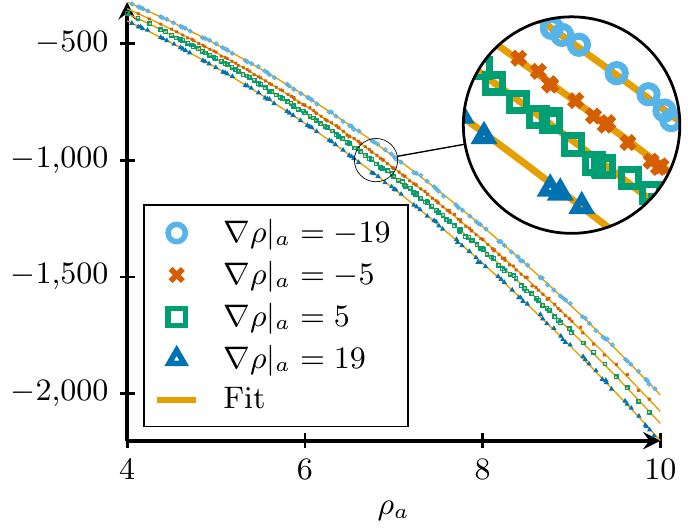}\fi
         \renewcommand{\tikzpicturename}{fig-matrix-entries-no-constraint-a-Equal-bPlus1}
         \iftikzjz \tikzsetnextfilename{\tikzpicturename}
         \input{\pathname\tikzpicturename.tikz}
         \else\includegraphics[scale=.57]{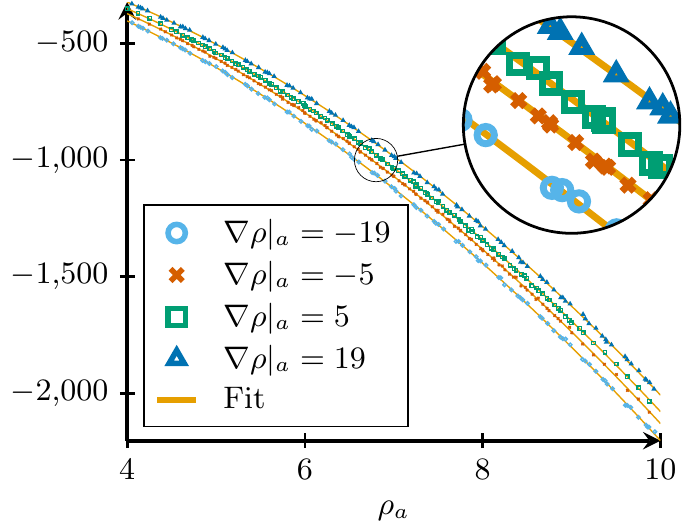}\fi
     } 
     \caption{Data set used in the unconstrained setting. Top left: Discrete set $V_\discr$ of pairs $(\rho_a, \nabla \rho|_a)$
       used to evaluate the discretized operator $\K_\rho$. Remaining plots: Shown are the matrix entries $(b,a)$ of
       $\langle \K_{\left(\rho_a+ \nabla \rho|_a ( x-x_a)\right)}\gamma_a,\gamma_b \rangle$ as function of $\rho_a$ and
       $\nabla \rho|_a$ for the symmetric zero-range process. The plots are for $a=b$ (top right), $a=b-1$ (bottom left) and
       $a=b+1$ (bottom right). For $x$, $x_a$ and the profiles $\gamma_a$ see Fig.~\ref{Fig:ProfilesSimulated}.}
     \label{Fig:MetricEntries}
\end{figure}

We remark that in the absence of the mass conservation constraint, good results can still be achieved as shown in
Fig.~\ref{Fig:PDE_results} (see also the corresponding movies Movie 3 and 4 in \ifarxiv \cite{Li2018a-movies}\else the
Supplemental Material\fi), although small deviations may be observed at large times (see left panel).  The discrete set $V_\discr$
used in these simulations is shown in Fig.~\ref{Fig:MetricEntries}, together with the entries for the operator and their
(independent) quadratic fit. To achieve the observed accuracy, a relatively large size of the set $V_\discr$ is
required. Specifically, $V_\discr$ consists of $6111$ points, in contrast to $228$ points when mass conservation is imposed.

\section{Covariance of the fluctuations}
\label{sec:covariance}

We now provide a proof of the relation
\ifarxiv
\begin{equation}
  \langle \K_z\gamma_a,\gamma_b \rangle
  =\lim_{h \searrow 0} \frac{1}{2h}\mathbb{E}\Big[ \left(Y_{\gamma_a}(t_0+h)- Y_{\gamma_a}(t_0)\right) \Bigr.  
   \Bigl. \cdot \left(Y_{\gamma_b}(t_0+h)- Y_{\gamma_b}(t_0) \right)\Big]
  \label{Eq:Covariance}
\end{equation}
\else
\begin{multline}
  \langle \K_z\gamma_a,\gamma_b \rangle
  =\lim_{h \searrow 0} \frac{1}{2h}\mathbb{E}\Big[ \left(Y_{\gamma_a}(t_0+h)- Y_{\gamma_a}(t_0)\right) \Bigr.  
   \\  
   \Bigl. \cdot \left(Y_{\gamma_b}(t_0+h)- Y_{\gamma_b}(t_0) \right)\Big]
  \label{Eq:Covariance}
\end{multline}
\fi used to compute the discretized operator from the rescaled local fluctuations $Y_{\gamma}$, defined as the stochastic limit of
$Y_{\gamma}^{\epsilon}=\frac{1}{\sqrt{\epsilon}}\langle z_{\epsilon}-z,\gamma \rangle$ as $\epsilon\rightarrow 0$. The proof
follows a similar argument to that in~\cite{Embacher2018a}, where the variation of the fluctuations for a specific operator (the
Wasserstein operator) is computed.

From the equations for $z_{\epsilon}$ and $z=\E[z_{\epsilon}]$, cf.~\eqref{eq:GF} and~\eqref{eq:stoch-GF} of the article, the
rescaled fluctuations follow
\begin{equation}
  \dd Y_{\gamma}= \langle {\mathcal M}_z Y, \gamma \rangle\ \dd t + \langle \sqrt{2\K_{z}}\dd W_{x,t},\gamma\rangle,
\end{equation}
where $Y$ is the limit of $(z_{\epsilon}-z)/ \sqrt{\epsilon}$ and ${\mathcal M}_z$ is a linear operator acting on $Y$, depending
on $z$, such that ${\mathcal M}_z Y$ is the limit of
$ \left(\K_{z_{\epsilon}}DS(z_{\epsilon})-\K_z DS(z) \right)/ \sqrt{\epsilon}$.  Additionally, by It\^o's formula, the function
$F(X_1,X_2)=\left(X_1-Y_{\gamma_a}(t_0) \right)\left(X_2-Y_{\gamma_b}(t_0) \right)$ satisfies for $X_1=Y_{\gamma_a}(t)$ and
$X_2=Y_{\gamma_b}(t)$
\ifarxiv
\begin{equation}
  \dd F=\left(Y_{\gamma_b}(t)-Y_{\gamma_b}(t_0) \right)\dd Y_{\gamma_a}
    + \left(Y_{\gamma_a}(t)-Y_{\gamma_a}(t_0) \right)\dd Y_{\gamma_b} + \langle 2\K_z\gamma_a,\gamma_b\rangle \df t.
\end{equation}
\else
\begin{multline}
  \dd F=\left(Y_{\gamma_b}(t)-Y_{\gamma_b}(t_0) \right)\dd Y_{\gamma_a}
   \\
    + \left(Y_{\gamma_a}(t)-Y_{\gamma_a}(t_0) \right)\dd Y_{\gamma_b} + \langle 2\K_z\gamma_a,\gamma_b\rangle \df t.
\end{multline}
\fi Then, the expectation appearing on the right-hand side of~\eqref{Eq:Covariance}, can be written as
\begin{equation}
  \label{Eq:proof1}
  \begin{split}
    \E&\left[ \left(Y_{\gamma_a}(t_0+h)- Y_{\gamma_a}(t_0)\right) \left(Y_{\gamma_b}(t_0+h)- Y_{\gamma_b}(t_0) \right)\right]\\
    &=\E\left[F(Y_{\gamma_a}(t_0+h),Y_{\gamma_b}(t_0+h))\right] =\E\left[\int_{t_0}^{t_0+h} \dd F \right]\\
    &=\E\left[ \int_{t_0}^{t_0+h}\left(Y_{\gamma_b}(t)-Y_{\gamma_b}(t_0) \right)\dd Y_{\gamma_a} \right] \\
    &\qquad+\E\left[ \int_{t_0}^{t_0+h}\left(Y_{\gamma_a}(t)-Y_{\gamma_a}(t_0) \right)\dd Y_{\gamma_b} \right] \\
    &\qquad+\E\left[\int_{t_0}^{t_0+h} \langle 2\mathcal{K}_z\gamma_a,\gamma_b\rangle\ \dd t \right].
  \end{split}
\end{equation}
The sought-after result then immediately follows if
\begin{equation}
  \label{Eq:proof3}
  \begin{split}
    &\lim_{h \searrow 0} \frac{1}{2h}\E\left[ \int_{t_0}^{t_0+h}\left(Y_{\gamma_b}(t)-Y_{\gamma_b}(t_0) \right)\dd Y_{\gamma_a} \right]=0, \\
    &\lim_{h \searrow 0} \frac{1}{2h}\E\left[ \int_{t_0}^{t_0+h}\left(Y_{\gamma_a}(t)-Y_{\gamma_a}(t_0) \right)\dd Y_{\gamma_b} \right]=0,
  \end{split}
\end{equation}
which we proceed to show. 

By H\"older's and Young's inequality
\begin{equation}
  \label{Eq:proof2}
  \begin{split}
    \E&\left[ \int_{t_0}^{t_0+h}\left(Y_{\gamma_b}(t)-Y_{\gamma_b}(t_0) \right)\dd Y_{\gamma_a}\right]\\
    &=\E\left[ \int_{t_0}^{t_0+h}\left(Y_{\gamma_b}(t)-Y_{\gamma_b}(t_0) \right) \langle {\mathcal M}_z Y, \gamma_a \rangle\ \dd t\right]\\
    & \leq \sqrt{\E\left[ \int_{t_0}^{t_0+h}\left(Y_{\gamma_b}(t)-Y_{\gamma_b}(t_0) \right)^2 \dd t\right]}\\
    & \phantom{\leq}\ \cdot\sqrt{\E\left[ \int_{t_0}^{t_0+h} \left(\langle {\mathcal M}_z Y, \gamma_a \rangle \right)^2\dd t\right]}\\
    &\leq \frac{1}{2} \int_{t_0}^{t_0+h} \E \left[ \left(Y_{\gamma_b}(t)-Y_{\gamma_b}(t_0) \right)^2\right] \dd t \\
    &+ \frac{1}{2} \int_{t_0}^{t_0+h} \E \left[\left(\langle {\mathcal M}_z Y, \gamma_a \rangle \right)^2 \right] \dd t,
  \end{split}
\end{equation}
and analogously for $\E\left[ \int_{t_0}^{t_0+h}\left(Y_{\gamma_a}(t_0+h)-Y_{\gamma_a}(t_0) \right)\dd Y_{\gamma_b}
  \right]$. Next, we define the auxiliary variables
\begin{equation}
  \begin{split}
    &Z_j(t)=\int_{t_0}^t \E \left[ \left(Y_{\gamma_j}(s)-Y_{\gamma_j}(t_0) \right)^2\right] \dd s \quad\text{and}\\
    &R_j(t)=\int_{t_0}^t \E \left[ \left(\langle {\mathcal M}_z Y, \gamma_j \rangle \right)^2\right] \dd s + \int_{t_0}^t \langle 2\mathcal{K}_z\gamma_j,\gamma_j\rangle  \dd s\\
  \end{split}
\end{equation}
with $j=a$ or $b$. From~\eqref{Eq:proof1}, \eqref{Eq:proof3} and~\eqref{Eq:proof2} for $\gamma_a=\gamma_b=\gamma_j$, it follows
that $\dot{Z}_j(t) \leq Z_j(t) + R_j(t)$, with $R_j(t)$ continuous. By Gronwall's lemma
\begin{equation}
  Z_j(t) \leq e^{(t-t_0)}\int_{t_0}^t e^{-(s-t_0)}R_j(s) \dd s,
\end{equation}
and, since $R_j(h)=\mathcal{O}(h)$, $Z_j(t_0+h)=\mathcal{O}(h^2)$. Then, using the second but last inequality
in~\eqref{Eq:proof2},
\begin{equation}
  \E\left[ \int_{t_0}^{t_0+h}\left(Y_{\gamma_b}(t)-Y_{\gamma_b}(t_0) \right)\dd Y_{\gamma_a}\right] = \mathcal{O}(h^{3/2}).
\end{equation}
Similarly, 
\begin{equation}
  \E\left[ \int_{t_0}^{t_0+h}\left(Y_{\gamma_a}(t)-Y_{\gamma_a}(t_0) \right)\dd Y_{\gamma_b}\right] = \mathcal{O}(h^{3/2}),
\end{equation}
which leads to~\eqref{Eq:proof3}, concluding the proof.

\vspace{0.1cm}




\def\cprime{$'$} \def\cprime{$'$} \def\cprime{$'$}
  \def\polhk#1{\setbox0=\hbox{#1}{\ooalign{\hidewidth
  \lower1.5ex\hbox{`}\hidewidth\crcr\unhbox0}}} \def\cprime{$'$}
  \def\cprime{$'$}

\end{document}